\documentclass[11pt]{article}
\usepackage[dvipdfmx]{color}
\usepackage{graphicx}
\usepackage{xcolor}
  \usepackage{amsthm,amsfonts}
  \usepackage{amsmath}
\usepackage{mathabx}
\usepackage{slashed}
\usepackage{ulem}
\usepackage{hyperref}
\usepackage{cleveref}
\usepackage{cite}

\newcommand{\bea}   {\begin{eqnarray}}
\newcommand{\eea}   {\end{eqnarray}}
\def\zzg{${\mathbb Z}_2\times{\mathbb Z}_2$-graded }

\begin{document}
\renewcommand{\thefootnote}{\fnsymbol{footnote}}

\thispagestyle{empty}

\title{Inequivalent quantizations from \\ gradings and ${\mathbb Z}_2\times {\mathbb Z}_2$  parabosons}
\author{ Francesco Toppan\thanks{{E-mail: {\it toppan@cbpf.br}}}
\\
\\
}
\maketitle

{\centerline{
{\it CBPF, Rua Dr. Xavier Sigaud 150, Urca,}}\centerline{\it{
cep 22290-180, Rio de Janeiro (RJ), Brazil.}}
~\\
\maketitle
\begin{abstract}
This paper introduces the parastatistics  induced by {\it \zzg algebras}. It  accommodates four kinds of particles:  ordinary bosons and three types of parabosons which mutually anticommute when belonging to different type
(so far, in the literature, only parastatistics induced by {\it \zzg superalgebras} and  producing parafermions have been considered).\par
It is shown how to detect \zzg parabosons in the multi-particle sector of a quantum model.
The difference with respect to a system composed by ordinary bosons is spotted by measuring some selected observables on certain given eigenstates. The construction
of the multi-particle states is made through the appropriate braided tensor product.\par
The application of ${\mathbb Z}_2$- and ${\mathbb Z}_2\times {\mathbb Z}_2$- gradings produces $9$ inequivalent multi-particle Hilbert spaces of a $4\times 4$ matrix oscillator.
The \zzg parabosonic Hilbert space is one of them.
\end{abstract}
\vfill
\rightline{CBPF-NF-002/21}
\newpage

\section{Introduction}

This paper introduces the parastatistics induced by \zzg Lie {\it algebras} and gives the proof that the associated
particles, the \zzg parabosons, can be detected by performing a measurement in the multi-particle sector of a quantum model.\par
\zzg Lie algebras and Lie superalgebras were introduced by Rittenberg and Wyler in \cite{{riwy1},{riwy2}}. The term ``{\it color
(super)algebra}"  was used (see also \cite{sch1}) to describe both cases.
\par
The particles (bosons and fermions) of an ordinary theory can be associated with $1$ bit of information (let's say $0$ for bosons and $1$ for fermions), while the particles of a \zzg theory are described by $2$ bits of information ($00$, $10$, $11$, $01$).\par
 The four types of particles in models based on \zzg Lie {\it superalgebras} are (see \cite{akt1}) the
ordinary bosons ($00$), two types  ($10$ and $01$) of parafermions and the exotic bosons ($11$); the parafermions of different type commute, while the exotic bosons anticommute with the parafermions.\par
 The four types of particles in models based on \zzg Lie {\it algebras} are (see \cite{kuto1}) the
ordinary bosons ($00$) and three types  ($10$, $11$ and $01$) of parabosons; the parabosons of different type anticommute.\par
The \zzg physics, with particles accommodated according to Lie superalgebras, is an obvious extension of ordinary physics. Indeed, ordinary bosons and fermions can be recovered from, respectively, the $00$ and $10$ sectors, while leaving empty the $11$ and $01$ sectors. Only recently, however, the open question that was lingering around was solved in \cite{top1}, by showing that the {\it colored world} of \zzg  Lie superalgebras produces quantum models which cannot be mimicked by black/white ordinary bosons and fermions alone. \par
Symbolically, this result can be expressed as
\bea
{\mathbb Z}_2^1\cdot{\boldsymbol {LSA}} &\subset & {\mathbb Z}_2^2\cdot{\boldsymbol {LSA}},
\eea 
meaning that the systems recovered from ${\mathbb Z}_2$-graded Lie superalgebras are a proper subset of those
recovered from ${\mathbb Z}_2\times {\mathbb Z}_2$-graded Lie superalgebras.\par
The approach of \cite{top1} emphasizes the role of the braided tensor product, as defined in  \cite{maj1}, in the construction of the multi-particle states. This approach is here extended to derive the physics of the parabosons obtained from \zzg Lie algebras. A symbolical consequence of the present work can be expressed as
\bea
{\mathbb Z}_2^0\cdot{\boldsymbol {LA}} &\subset & {\mathbb Z}_2^2\cdot{\boldsymbol {LA}},
\eea 
meaning that the bosonic systems recovered from ordinary Lie algebras are a proper subset of those recovered
from ${\mathbb Z}_2\times {\mathbb Z}_2$-graded Lie algebras.
\par
In the first years after the introduction of \zzg Lie superalgebras their physical applications received some limited attention, see \cite{{luri1},{vas1},{jyw1},{zhe1}}. More recently, a systematic investigation of their role as symmetries of dynamical systems 
started. Indeed, they appear \cite{{aktt1},{aktt2}} as symmetries of the L\'evy-Leblond equations for nonrelativistic spinors; furthermore, 
classical invariant worldline \cite{akt1} and two-dimensional \cite{bru1} sigma models have been constructed, invariant quantum mechanical models have been presented in \cite{{brdu1},{akt2}} and conformal quantum mechanics in \cite{aad1}. The parastatistics induced by \zzg Lie superalgebras was introduced in \cite{{yaji1},{yaji2}} and further investigated in
\cite{{kada1},{kaha1},{kaha2},{kan1}, {tol1}, {stvj1}, {top1}}. \par
Despite this activity on graded Lie superalgebras, the possibility of applications of \zzg Lie algebras has been  ignored, probably because they are not extensions of ordinary \cite{kac1} Lie superalgebras and do not include fermions.
Indeed, a consistent number of present works are focusing on even larger (the ${\mathbb Z}_2^n$, for $n>2$) graded extension of Lie superalgebras, see e.g. \cite{{bru2},{aad2},{doai1}} and references therein for the mathematical literature. \par
On the other hand, as pointed out very recently in \cite{kuto1}, several costructions (invariant models, the graded superspace of \cite{brdu2}, etc.) which are available for graded superalgebras can be extended to
\zzg Lie algebras.
This is the starting point of the present investigation.\par
The scheme of the paper is as follows. It is shown at first that the $4 \times 4 $ matrix Hamiltonian discussed in
\cite{{brdu1},{akt2}}, besides being supersymmetric and invariant under the one-dimensional \zzg Poincar\'e superalgebra, is also invariant under a \zzg Lie algebra. This offers the possibility to apply the \zzg parabosonic statistics to its multi-particle sector. \par
The Hilbert space is constructed for the special case of the harmonic oscillator potential. It is shown that different statistics can be implemented by using ${\mathbb Z}_2$- and ${\mathbb Z}_2\times{\mathbb Z}_2$- gradings. As a consequence, a total number of $9$ inequivalent multi-particle Hilbert spaces are encountered. This statement can also be rephrased as ``$9$ inequivalent multi-particle quantizations". The analysis of \cite{top1} discussed just three of them (bosonic, supersymmetric and \zzg parafermionic variants). 
Among the extra quantizations presented in this paper a case  corresponds to
the \zzg parabosons, while another case corresponds to a different implementation of the \zzg parafermions.\par
The multi-particle states are constructed by taking the raising operators as elements of a Universal Enveloping graded Lie (super)algebra and by applying the associated coproducts.   \par
The proof that the $9$ variants indeed produce inequivalent multi-particle models is given. In particular, observables discriminating the \zzg parabosonic variant from the bosonic one are constructed.\par
More comments about the results and the future perspectives are given in the Conclusions.
~\par
The index of the paper is:\par
{\bf 1} - {Introduction},\par
{\bf 2} - {The $4\times 4$ graded Hamiltonian},\par
{\bf 3} - {The construction of multi-particle Hilbert spaces},\par
{\bf 4} - {The $9$ inequivalent $2$-particle Hilbert spaces},\par
{\bf 5} - {Discriminating $2$-particle observables},\par
{\bf 6} - {Conclusions},\par
{Appendix {\bf A} - {Relevant formulas for graded (super)algebras},\par
{Appendix {\bf B} - {Representations of $2$-particle observables}.

\section{The ${\boldsymbol{4\times 4}}$ graded Hamiltonian}

The $4\times 4$ hermitian matrix Hamiltonian $H$, given by
\bea\label{4x4ham}
H&=& {\footnotesize{\frac{1}{2}\left(
\begin{array}{cccc} -\partial_x^2+W^2(x)+W'(x)&0&0&0\\0&-\partial_x^2+W^2(x)+W'(x)&0&0\\0&0&-\partial_x^2+W^2(x)-W'(x)&0\\0&0&0&-\partial_x^2+W^2(x)-W'(x)
\end{array}
\right)}},\nonumber\\&&
\eea
depends on the prepotential $W(x)$. We set  in the above formula $W'(x)=\frac{d}{dx}W(x)$.\\
$H$ is invariant, see \cite{{brdu1},{akt2}}, under both supersymmetry and the one-dimensional \zzg Poincar\'e superalgebra.\par
We point out here that $H$ is also invariant under a \zzg Lie algebra. Let us introduce the hermitian first-order matrix operators $Q_{10}$, $Q_{01}$ and constant matrix $Z$ as
\bea\label{qqz}
{Q}_{10}&=&{\footnotesize{\frac{-i}{\sqrt{2}}\left(
\begin{array}{cccc} 0&0&\partial_x+W(x)&0\\0&0&0&\partial_x+W(x)\\\partial_x-W(x)&0&0&0\\0&\partial_x-W(x)&0&0
\end{array}
\right)}},\nonumber\\
{ Q}_{01}&=&{\footnotesize{\frac{1}{\sqrt{2}}\left(
\begin{array}{cccc} 0&0&0&\partial_x+W(x)\\0&0&\partial_x+W(x)&0\\0&-\partial_x+W(x)&0&0\\-\partial_x+W(x)&0&0&0
\end{array}
\right)}},\nonumber\\
Z&=&{\footnotesize{\left(
\begin{array}{cccc} 0&1&0&0\\1&0&0&0\\0&0&0&-1\\0&0&-1&0
\end{array}
\right).}}
\eea
~\par
The Hamiltonian $H$ is invariant under the \zzg abelian Lie algebra ${\mathfrak a}$ defined by the following set of (all vanishing) $6$  (anti)commutators:
\bea\label{gradedalginv}
[H,Q_{10}]=[H,Q_{01}]=[H,Z]=0, &&\{Q_{10},Q_{01}\}=\{Z,Q_{10}\}=\{Z,Q_{01}\}=0.
\eea
The algebra ${\mathfrak a}$ is listed as ``$A7$" in the Table {\bf 1} classification  of minimal graded algebras presented in\cite{kuto1}. The grading assignment, according to the (\ref{gradedmatrices}) decomposition, of the ${\mathfrak a}$ generators is
\bea
&H\in {\mathfrak a}_{00}, \quad Q_{10}\in{\mathfrak a}_{10},\quad 
Q_{01}\in {\mathfrak a}_{01},\quad Z\in {\mathfrak a}_{11}.&
\eea

The operators $Q_{10}$, $Q_{01}$ are square roots of the Hamiltonian ($Q_{10}^2=Q_{01}^2=H$).  It follows that, besides (\ref{gradedalginv}), $H$ is invariant under the ${\mathbb Z}_2$-graded Lie superalgebra
\bea\label{susyham}
&\{Q_{10},Q_{10}\}=\{Q_{01},Q_{01}\}= 2H, \qquad \{Q_{10},Q_{01}\}=0,\qquad [H,Q_{10}]=[H,Q_{01}]=0,&
\eea
which defines $H$ as a supersymmetric quantum mechanics \cite{wit1} Hamiltonian. \par We anticipate that these two
graded invariant structures produce inequivalent quantum models in the multi-particle sectors. Essentially, this is due to the fact that the fermions present in (\ref{susyham}) obey the Pauli exclusion principle; this is not the case for the parabosons that, as we will see, are obtained from (\ref{gradedalginv}).\par
If we specialize $W(x)=-x$, $H$ becomes the Hamiltonian of the one-dimensional, $4\times 4$ matrix oscillator. It will be denoted as $H_{osc}$; we have
\bea\label{hamosc}
H_{osc}&=& {\footnotesize{\frac{1}{2}\left(
\begin{array}{cccc} -\partial_x^2+x^2-1&0&0&0\\0&-\partial_x^2+x^2-1&0&0\\0&0&-\partial_x^2+x^2+1&0\\0&0&0&-\partial_x^2+x^2+1
\end{array}
\right).}}
\eea
The single-particle Hilbert space ${\cal H}$ of the $H_{osc}$ Hamiltonian was constructed in \cite{{brdu1},{top1}}. It
is obtained by applying raising operators to a lowest weight vector state denoted as $|0;00\rangle$.\par
The creation/annihilation oscillators $a, a^\dagger$, given by
\bea\label{aadagger}
a=\frac{i}{\sqrt{2}}(\partial_x+x), && a^\dagger =\frac{i}{\sqrt{2}}(\partial_x-x),
\eea
satisfy the commutator 
\bea
[a,a^\dagger] &=& 1.
\eea
The matrix raising (lowering) operators $f_{11}^\dagger, f_{10}^\dagger, f_{01}^\dagger$ ($f_{11}, f_{10}, f_{01}$) can be introduced as
\bea\label{fffmatrices}
&f_{11}^\dagger= {\footnotesize{\left(
\begin{array}{cccc} 0&0&0&0\\1&0&0&0\\0&0&0&0\\0&0&0&0
\end{array}
\right),}}\quad f_{10}^\dagger= {\footnotesize{\left(
\begin{array}{cccc} 0&0&0&0\\0&0&0&0\\1&0&0&0\\0&0&0&0
\end{array}
\right),}}\quad f_{01}^\dagger= {\footnotesize{\left(
\begin{array}{cccc} 0&0&0&0\\0&0&0&0\\0&0&0&0\\1&0&0&0
\end{array}
\right),}}&\nonumber\\
&f_{11}= {\footnotesize{\left(
\begin{array}{cccc} 0&1&0&0\\0&0&0&0\\0&0&0&0\\0&0&0&0
\end{array}
\right),}}\quad f_{10}= {\footnotesize{\left(
\begin{array}{cccc} 0&0&1&0\\0&0&0&0\\0&0&0&0\\0&0&0&0
\end{array}
\right),}}\quad f_{01}= {\footnotesize{\left(
\begin{array}{cccc} 0&0&0&1\\0&0&0&0\\0&0&0&0\\0&0&0&0
\end{array}
\right).}}&
\eea
The suffix is chosen in order to denote, in the ${\mathbb Z}_2\times{\mathbb Z}_2$-gradings, the matrix decompositions expressed in (\ref{gradedmatrices}).\par
In terms of these operators the Hamiltonian $H_{osc}$ can be re-expressed as
\bea\label{hosc2}
H_{osc} &=& a^\dagger a\cdot{\mathbb I}_4 + f_{10}^\dagger f_{10} +f_{01}^\dagger f_{01}=a^\dagger a\cdot{\mathbb I}_4 +\Lambda,\quad
{\textrm{with}}\quad\Lambda=diag(0,0,1,1).
\eea
Here and in the following we denote a $m\times m$ identity matrix as ${\mathbb I}_m$.\par
The normalized lowest weight vector $|0;00\rangle$ satisfies the conditions
\bea
&a|0;00\rangle=f_{11}|0;00\rangle=f_{10}|0;00\rangle=f_{01}|0;00\rangle=0.&
\eea
We have
\bea\label{lwv000}
|0;00\rangle&=&\pi^{-\frac{1}{4}}e^{-\frac{1}{2}x^2}  {\footnotesize{\left(
\begin{array}{c} 1\\0\\0\\0
\end{array}
\right).}}
\eea
The single-particle Hilbert space ${\cal H}$ is spanned by the orthonormal vectors \\$|n;00\rangle,|n;11\rangle,|n;10\rangle,|n;01\rangle$ introduced through
\bea\label{spanningvectors}
&\begin{array}{ll}
|n;00\rangle =\frac{{(a^\dagger})^n}{\sqrt{n!}}|0;00\rangle, \qquad&
|n;10\rangle =\frac{{(a^\dagger})^n}{\sqrt{n!}}f_{10}^\dagger|0;00\rangle, \\
|n;11\rangle =\frac{{(a^\dagger})^n}{\sqrt{n!}}f_{11}^\dagger|0;00\rangle, \qquad&
|n;01\rangle =\frac{{(a^\dagger})^n}{\sqrt{n!}}f_{01}^\dagger|0;00\rangle.
\end{array}&
\eea
At most a single power of $f_{11}^\dagger,f_{10}^\dagger,f_{01}^\dagger$ enters the spanning vectors since we have,
for any pair of such operators,
\bea
f_\sharp^\dagger f_\flat^\dagger &=& 0,\qquad {\textrm{with}}\quad \sharp,\flat\in\{11,10,01\}.
\eea
Due to the commutators
\bea\label{commut}
&[H_{osc},a^\dagger]=a^\dagger, \quad [H_{osc},f_{10}^\dagger]=f_{10}^\dagger,\quad
[H_{osc},f_{01}^\dagger]=f_{01}^\dagger,\quad [H_{osc},f_{11}^\dagger]=0,&
\eea
the (\ref{spanningvectors}) states are energy eigenstates whose eigenvalues are read from
\bea
&\begin{array}{ll}
H_{osc}|n;00\rangle=n|n;00\rangle,\qquad&
H_{osc}|n;10\rangle=(n+1)|n;10\rangle,\\
H_{osc}|n;11\rangle=n|n;11\rangle,\qquad &
H_{osc}|n;01\rangle=(n+1)|n;01\rangle.
\end{array}
&
\eea
One should note that the vacuum state is doubly degenerate:
\bea
H_{osc}|0;00\rangle=H_{osc}|0;11\rangle &=&0.
\eea
For later convenience we introduce the exchange matrices $X_{11}, X_{10}, X_{01}$.  They are hermitian operators which mutually interchange the $11$, $10$ and $01$ sectors. Their suffix indicates the (\ref{gradedmatrices}) decomposition when
a ${\mathbb Z}_2\times{\mathbb Z}_2$-grading is applied. We have
\bea\label{xxxmatrices}
&X_{11}= {\footnotesize{\left(
\begin{array}{cccc} 0&0&0&0\\0&0&0&0\\0&0&0&1\\0&0&1&0
\end{array}
\right),}}\quad X_{10}= {\footnotesize{\left(
\begin{array}{cccc} 0&0&0&0\\0&0&0&1\\0&0&0&0\\0&1&0&0
\end{array}
\right),}}\quad X_{01}= {\footnotesize{\left(
\begin{array}{cccc} 0&0&0&0\\0&0&1&0\\0&1&0&0\\0&0&0&0
\end{array}
\right).}}&
\eea
The matrices $X_{11}, X_{10}, X_{01}$ are the building blocks in the construction of the observables which are presented in Section {\bf 5}.


 \section{The construction of multi-particle Hilbert spaces}

The $n$-particle Hilbert space ${\cal H}^{(n)}$ of the (\ref{hamosc}) Hamiltonian $H_{osc}$ is a subset
of the tensor products of $n$ single-particle Hilbert spaces ${\cal H}$:
\bea
{\cal H}^{(n)} &\subset {\cal H}^{\otimes n}.
\eea
${\cal H}^{(n)}$ is a lowest weight vector space whose lowest weight vector $|0;00\rangle^{(n)}$ is a tensor
product of the single-particle lowest weight vector $|0;00\rangle$ given in (\ref{lwv000}):
\bea
|0;00\rangle^{(n)} &=&|0;00\rangle^{\otimes n}.
\eea
The space coordinates entering the tensor products of the Hilbert spaces ${\cal H}^{(n)}$ are denoted as
$x_1,x_2,\ldots, x_n$. In the $2$-particle case we set, for simplicity, $x_1=x$, $x_2=y$. Therefore, the
normalized lowest weight vector $|0;00\rangle^{(2)}$ is
\bea\label{lwv2}
|0;00\rangle^{(2)}&=&\pi^{-\frac{1}{2}}e^{-\frac{1}{2}(x^2+y^2)} \cdot v_1.
\eea
Here and in the following we denote as $v_j$, for $j=1,2,\ldots, 16$, the $16$-component column vector with entry $1$ in the $j$-th position and $0$ otherwise.\par
The construction of the $2$-particle Hilbert space assumes the raising operators $a^\dagger, f_{11}^\dagger,f_{10}^\dagger,f_{01}^\dagger$ introduced in (\ref{aadagger},\ref{fffmatrices}) to be elements of a graded algebra ${\mathfrak g}$ (the admissible gradings for ${\mathfrak g}$ are discussed in Section ${\bf 4}$). The graded algebra ${\mathfrak g}$ defines its Universal Enveloping Algebra $U\equiv {\cal U}({\mathfrak g})$. As recalled in Appendix {\bf A},
$U$ is endowed with a Hopf algebra structure and in particular of an operation, the coproduct $\Delta$, which satisfies (\ref{coproduct},\ref{coassoc},\ref{coproductprop}).\par
The $2$-particle states are recovered from applying the coproducts
\bea
\Delta \left( ( {a^\dagger})^n)(f_{11}^\dagger)^{r_{11}}(f_{10}^\dagger)^{r_{10}}(f_{01}^\dagger)^{r_{01}}\right)&\in& U\otimes U
\eea
to the vector $
|0;00\rangle^{(2)}$ which induces the lowest weight representation. Following the convention of Appendix {\bf A}, a hat denotes the evaluation of the coproduct in the given representation. Therefore
\bea
{\widehat {\Delta \left( ( {a^\dagger})^n)(f_{11}^\dagger)^{r_{11}}(f_{10}^\dagger)^{r_{10}}(f_{01}^\dagger)^{r_{01}}\right)}}&\in& End({\cal H}^{(2)}).
\eea
The Hilbert space ${\cal H}^{(2)}$ is spanned by
\bea\label{hilbert2span}
|n;r_{11},r_{10},r_{01}\rangle^{(2)}&=&{\widehat {\Delta \left( ( {a^\dagger})^n)(f_{11}^\dagger)^{r_{11}}(f_{10}^\dagger)^{r_{10}}(f_{01}^\dagger)^{r_{01}}\right)}}\cdot
|0;00\rangle^{(2)}.
\eea
The identification $
|0;0,0,0\rangle^{(2)}\equiv 
|0;00\rangle^{(2)} $ holds.\par
In (\ref{hilbert2span}) $n$ is a non-negative integer ($n\in {\mathbb N}_0$); the restrictions on $r_{11},r_{10}, r_{01}$, as discussed in Section {\bf 4}, depend on the grading.\par 
As a useful example, it follows that the formula of the $2$-particle creation operator ${\widehat{\Delta(a^\dagger})}$ is
\bea\label{deltaadagger}
 {\widehat{\Delta(a^\dagger})}&=&\frac{i}{\sqrt{2}}(\partial_x-x+\partial_y-y).
\eea
By enlarging the graded algebra ${\mathfrak{g}}$ and its induced Universal Enveloping Algebra with the addition of observables such as $H_{osc}$, we can determine their actions on the multi-particle sectors. The $2$-particle Hamiltonian $H_{osc}^{(2)}$ reads as
\bea
H_{osc}^{(2)}&=&  {\widehat{\Delta(H_{osc}})}= H_{osc}\otimes {\mathbb I}_4+{\mathbb I}_4\times H_{osc}.
\eea
The construction of the $n+1$-particle Hilbert spaces, for $n>1$, is made iteratively by replacing $\Delta\equiv \Delta^{(1)}$ with $\Delta^{(n)}$. Induced by the coassociativity (\ref{coassoc}) of the coproduct, $\Delta^{(n)}$ is
defined as
\bea
\Delta^{(n)} &=& (id\otimes\Delta^{(1)} ) \Delta^{(n-1)}, \qquad (\Delta^{(1)}\equiv \Delta).
\eea

\section{The ${\boldsymbol{9}}$ inequivalent ${\boldsymbol{2}}$-particle Hilbert spaces}

The oscillator Hamiltonian $H_{osc}$ given in (\ref{hamosc}) possesses nine inequivalent multi-particle quantizations.
They are induced by the different gradings assigned to the raising operators $f_{11}^\dagger, f_{10}^\dagger, f_{01}^\dagger$ introduced in (\ref{fffmatrices}) and under the assumption that the lowest weight vector is bosonic. Three of the quantizations (bosonic, supersymmetric and a version of the \zzg parafermions) were already discussed in
\cite{top1}. The extra quantizations are divided into {\it standard} and {\it non-standard}.  Among the standard ones we obtain the \zzg parabosons; the non-standard ones include an alternative quantization based on \zzg parafermions.\par
The construction goes as follows: 
in one case (the bosonic one) $f_{11}^\dagger, f_{10}^\dagger, f_{01}^\dagger$ are  assumed to be elements of an ordinary abelian Lie algebra; alternatively,  they are assumed to be even/odd elements of a ${\mathbb Z}_2$-graded abelian Lie superalgebra, of a \zzg abelian Lie superalgebra (parafermions) or of a
\zzg abelian Lie algebra (parabosons).\par
We proceed at first to discuss the $6$ standard gradings.

\subsection{The  ${\boldsymbol{6}}$ standard gradings}

In the ${\mathbb Z}_2$-grading assignment the $4\times 4$ matrix Hamiltonian $H_{osc}$ corresponds to a block-diagonal supermatrix of $(4-p|p)$ type, with $p=0,1,2,3$. The $(4|0)$ case for $p=0$ coincides with the ordinary bosonic matrix. The $p=4$ case is excluded if we require the vacuum state to be even (bosonic).\par
The six assignments are:\par
~\par
$1$) ~~~ $\{f_{11}^\dagger,f_{10}^\dagger,f_{01}^\dagger\}\in 0$, ~~$\{\emptyset\}\in 1$   ~~~ for $(4|0)$;\par
$2$) ~~~ $\{f_{11}^\dagger,f_{10}^\dagger\}\in 0$, ~~~~~~~$\{f_{01}^\dagger\}\in 1$  ~~~for $(3|1)$;\par
$3$) ~~~ $\{f_{11}^\dagger\}\in 0$, ~~~~~~~$\{f_{10}^\dagger, f_{01}^\dagger\}\in 1$ ~~~for $(2|2)$;\par
$4$) ~~~ $\{\emptyset\}\in 0$, ~~~$\{f_{11}^\dagger, f_{10}^\dagger, f_{01}^\dagger\}\in 1$ ~~ for $(1|3)$;\par 
$5$) ~~~ $\{f_{11}^\dagger,f_{10}^\dagger,f_{01}^\dagger\}\in {\mathbb Z}_2^2\cdot LSA$;\par
$6$) ~~~ $\{f_{11}^\dagger,f_{10}^\dagger,f_{01}^\dagger\}\in {\mathbb Z}_2^2\cdot LA$.\par
~\par
The corresponding vanishing (anti)commutators defining the graded abelian algebras ${\mathfrak{a}}_j$, where $j=1,2,\ldots, 6$, are\
\bea\label{standardgrading}
\relax {\mathfrak{a}}_1: && [f_{11}^\dagger,f_{10}^\dagger]=[f_{10}^\dagger,f_{01}^\dagger]=[f_{01}^\dagger,f_{11}^\dagger]=0;\nonumber\\
\relax {\mathfrak{a}}_{2}: && [f_{11}^\dagger,f_{10}^\dagger]=[f_{10}^\dagger,f_{01}^\dagger]=[f_{01}^\dagger,f_{11}^\dagger]=\{f_{01}^\dagger,f_{01}^\dagger \}=0;\nonumber\\
\relax {\mathfrak{a}}_{3}: && [f_{11}^\dagger ,f_{10}^\dagger]=\{f_{10}^\dagger,f_{01}^\dagger\}=[f_{01}^\dagger,
f_{11}^\dagger]=\{f_{10}^\dagger,f_{10}^\dagger\}=\{f_{01}^\dagger,f_{01}^\dagger\}=0;\nonumber\\
\relax {\mathfrak{a}}_{4}: && \{f_{11}^\dagger,f_{10}^\dagger\}=\{f_{10}^\dagger,f_{01}^\dagger \}=\{f_{01}^\dagger,f_{11}^\dagger\}= \{f_{11}^\dagger,f_{11}^\dagger\}=\{f_{10}^\dagger,f_{10}^\dagger\}=\{f_{01}^\dagger,
f_{01}^\dagger\}=0;\nonumber\\
\relax {\mathfrak{a}}_5: && \{f_{11}^\dagger,f_{10}^\dagger\}=[f_{10}^\dagger,f_{01}^\dagger]=\{f_{01}^\dagger,f_{11}^\dagger\}=\{f_{10}^\dagger,f_{10}^\dagger\}=\{f_{01}^\dagger,f_{01}^\dagger\}=0;\nonumber\\
\relax {\mathfrak{a}}_{6}: && \{f_{11}^\dagger,f_{10}^\dagger\}=\{f_{10}^\dagger,f_{01}^\dagger\}=\{f_{01}^\dagger,f_{11}^\dagger \}=0.
\eea

The three cases already discussed in \cite{top1} correspond to the numbers $1$ (the bosonic version of the theory), $3$ (the supersymmetric version) and $5$ (a \zzg parafermionic version).\par
In the above construction we followed the standard block-diagonal matrix format of Lie superalgebras and, for the ${\mathbb Z}_2\times{\mathbb Z}_2$ grading, the (\ref{gradedmatrices}) decomposition. Non-standard supermatrix formats are discussed in \cite{dggt1}.  The procedure for the non-standard decompositions is presented in the subsection {\bf 4.2}. \par 
Each algebra ${\mathfrak a}_j$ is extended to the graded algebra ${\overline{\mathfrak{a}}}_j=\{a^\dagger,f_{11}^\dagger,f_{10}^\dagger,f_{01}^\dagger\}$ which contains the creation operator $a^\dagger$, introduced in (\ref{aadagger}), as extra generator. Depending on the case, $a^\dagger $ belongs to either the $0$- or the $00$-sector. Its commutators are vanishing
($[a^\dagger, f_\sharp^\dagger]=0$ for $\sharp=11,10,01$).\par
The multi-particle quantizations are recovered, as explained in Section {\bf 3}, from the coproducts defined  on the corresponding Universal Eveloping Algebras ${\cal U}({\overline{{\mathfrak{a}}}}_j)$.  The multi-particle states are constructed according to formula (\ref{hilbert2span}).
The signs entering the braided tensor products depend on the different grading assignments of each one of the above cases. They are given by (\ref{innerproducts},\ref{braidedtensor}).\par
The restrictions on the $r_{11}, r_{10}, r_{01}$ exponents entering (\ref{hilbert2span}) are due to these respective signs. For instance, in the parafermionic quantization $r_{10}$ takes the values $0,1$; the values taken by $r_{10}$ in the parabosonic case are $0,1,2$.\par
The $6$ standard multi-particle quantizations, associated to the respective (\ref{standardgrading}) gradings, are denoted as 
follows:
\bea \begin{array}{ccc}\label{standardquant}
 {\boldsymbol{(4|0)}}: \quad{\mathfrak{a}}_1,\qquad&{\boldsymbol{(2|2)}}: \quad{\mathfrak{a}}_3,\qquad&
{~~\boldsymbol{{\mathbb Z}_2^2${\bf -PF}$}}: \quad{\mathfrak{a}}_5,\\
 {\boldsymbol{(3|1)}}: \quad{\mathfrak{a}}_2,\qquad&
 {\boldsymbol{(1|3)}}: \quad{\mathfrak{a}}_4,\qquad&
{~~\boldsymbol{{\mathbb Z}_2^2${\bf -PB}$}}: \quad{\mathfrak{a}}_6.
\end{array}
\eea
In the last column {\bf PF} and {\bf PB} stand for, respectively, parafermions and parabosons.

\subsection{The  ${\boldsymbol{3}}$ non-standard gradings}

The non-standard cases are obtained by applying decompositions of the supermatrices which do not coincide with
the ordinary block-diagonal decompositions; these non-standard formats are discussed in \cite{dggt1}.
For the model under consideration these extra cases can be recovered from standard  decompositions applied to a different diagonal Hamiltonian whose diagonal entries are permuted with respect to $H_{osc}$.  \par
Before proceeding with the construction of the non-standard quantizations let us recall that the (\ref{fffmatrices}) raising operators $f_{11}^\dagger,f_{10}^\dagger, f_{01}^\dagger$ create, see (\ref{commut}), particles of respective energy $0,1,1$.\par
In a ${\mathbb Z}_2$-grading, the standard decomposition of a vector $v^T=(B,B,F,F)$ with $2$ bosons and $2$ fermions can be replaced, for instance, by the decomposition $v^T=(B,F,B,F)$. In these two examples the entries of the fermionic supermatrices are respectively accommodated according to
{\footnotesize{\bea
{\textrm{\large standard case:}}\quad \left(\begin{array}{cccc}
0&0&\ast&\ast\\
0&0&\ast&\ast\\
\ast&\ast&0&0\\
\ast&\ast&0&0
\end{array}
\right),&&
{\textrm{\large non-standard case:}}\quad \left(\begin{array}{cccc}
0&\ast&0&\ast\\
\ast&0&\ast&0\\
0&\ast&0&\ast\\
\ast&0&\ast&0
\end{array}
\right).
\eea
}}
For $3$ bosons and $1$ fermion we can pass from $v^T=(B,B,B,F)$ to, e.g., $v^T=(B,F,B,B)$.  In these new examples the entries of the fermionic supermatrices are respectively accommodated according to
{\footnotesize{\bea
{\textrm{\large standard case:}}\quad \left(\begin{array}{cccc}
0&0&0&\ast\\
0&0&0&\ast\\
0&0&0&\ast\\
\ast&\ast&\ast&0
\end{array}
\right),&&
{\textrm{\large non-standard case:}}\quad \left(\begin{array}{cccc}
0&\ast&0&0\\
\ast&0&\ast&\ast\\
0&\ast&0&0\\
0&\ast&0&0
\end{array}
\right).
\eea
}}
The key issue to notice is that the raising operator $f_{11}^\dagger$ becomes fermionic in the non-standard decompositions above. This implies that the Pauli exclusion principle applies to the $0$-energy particles created by
$f_{11}^\dagger$. In the standard cases these particles are bosons. This affects the degeneracy of the energy levels of the multi-particle Hamiltonian producing inequivalent results.\par
Similarly, a non-standard decomposition of a ${\mathbb Z}_2\times{\mathbb Z}_2$-graded Lie superalgebra is realized, e.g., by
accommodating the $00, 11,10,01$ sectors according to
{\small{\bea\label{gradedmatricesns}
M_{00}=\left(\begin{array}{cccc} \ast&0&0&0\\0&\ast&0&0\\0&0&\ast&0\\0&0&0&\ast\end{array}\right), && M_{10}=\left(\begin{array}{cccc} 0&\ast&0&0\\\ast&0&0&0\\0&0&0&\ast\\0&0&\ast&0\end{array}\right),\nonumber\\
M_{11}=\left(\begin{array}{cccc} 0&0&\ast&0\\0&0&0&\ast\\\ast&0&0&0\\0&\ast&0&0\end{array}\right), && M_{01}=\left(\begin{array}{cccc} 0&0&0&\ast\\0&0&\ast&0\\0&\ast&0&0\\\ast&0&0&0\end{array}\right).\eea
}}
Contrary to the standard decomposition (\ref{gradedmatrices}), in this case the $0$-energy particles created by $f_{11}^\dagger$ are no longer exotic bosons, but parafermions.\par 
On the other hand in the parabosonic case  induced by the  \zzg Lie algebra,  the non-standard decomposition above does not produce a inequivalent quantization with respect to the standard decomposition. This is so because, as already recalled, for parabosons the three sectors
$11$, $10$ and $01$ share the same properties and can be mutually interchanged.\par
A careful inspection shows that in three cases the non-standard decompositions for the Hamiltonian $H_{osc}$ are not equivalent to the standard ones. Nevertheless, in all three cases these decompositions can be recovered from their corresponding standard ones after changing the Hamiltonian $H_{osc}=a^\dagger a\cdot {\mathbb I}_4+\Lambda$, with $\Lambda=diag(0,0,1,1)$,  into the permuted Hamiltonian ${\overline H}_{osc}$ given by
\bea\label{overlinehamosc}
{\overline H}_{osc} &=& a^\dagger a\cdot {\mathbb I}_4 +{\overline\Lambda}, \quad{\textrm{with}}\quad  {\overline\Lambda}=diag(0,1,1,0).
\eea
These three non-standard multi-particle quantizations are denoted as ${\boldsymbol{(3|1)_{ns}}}$, ${\boldsymbol{(2|2)_{ns}}}$, $  ~{\boldsymbol{{\mathbb Z}_2^2${\bf -PF}$_{ns}}}$. Their corresponding graded algebras are
\bea \label{nsgradings}
 {\boldsymbol{(3|1)_{ns}}}: &&{\mathfrak{a}}_2\quad {\textrm{for}} \quad H_{osc}\mapsto {\overline H}_{osc},\nonumber\\
 {\boldsymbol{(2|2)_{ns}}}: &&{\mathfrak{a}}_3\quad {\textrm{for}} \quad H_{osc}\mapsto {\overline H}_{osc},\nonumber\\
{~~\boldsymbol{{\mathbb Z}_2^2${\bf -PF}$_{ns}}}: &&{\mathfrak{a}}_5\quad {\textrm{for}} \quad H_{osc}\mapsto {\overline H}_{osc}.
\eea

\subsection{The  ${\boldsymbol{2}}$-particle Hilbert spaces}

The orthonormal vectors spanning the $2$-particle Hilbert spaces, from  (\ref{lwv2},\ref{hilbert2span},\ref{deltaadagger}), have the form
\bea
|m; I\rangle &=& \frac{1}{\sqrt{ m!}}\left(\frac{i}{\sqrt{2}}(\partial_x+\partial_y-x-y)\right)^m\cdot(\pi^{-\frac{1}{2}}e^{-\frac{1}{2}(x^2+y^2)})\otimes V_I,
\eea 
where $V_I$ are $16$-component constant orthonormal vectors which can be expressed in the $v_j$ basis (we recall that $v_j$ has entry $1$ in the $j$-{th} position and $0$ otherwise).\par  
The $2$-particle Hilbert spaces  induced by the $6$ standard gradings will be denoted as
${\cal H}_{k}^{(2)}$; the suffix $k=1,2,\ldots,6$ denotes the respective (\ref{standardgrading}) graded algebras.
The finite dimensional Hilbert spaces ${\overline{\cal H}}_{k}^{(2)}\subset{\cal H}_k^{(2)}$ are spanned by the $V_I$ vectors by taking $m=0$ (the gaussian factor can be dropped for convenience).
The spanning vectors $V_I$ entering the six standard quantizations (\ref{standardquant}) are read from the following table:
\bea&\nonumber
\begin{array}{|l|c||c|c|c||c|c|}\hline  &{\boldsymbol{ (4|0)}}&{\boldsymbol{ (3|1)}}&{\bf (2|2)}&{\boldsymbol{ (1|3)}}&{{{\mathbb Z}_2^2}}${\bf -PF}$&{\mathbb Z}_2^2${\bf -PB}$ \\  \hline
V_1=v_1&$X$&$X$&$X$&$X$&$X$&$X$\\  \hline
V_2=v_6&$X$&$X$&$X$&&$X$&$X$\\  \hline 
V_3=v_{11}&$X$&$X$&&&&$X$\\ \hline 
V_4=v_{16}&$X$&&&$$&&$X$\\  \hline 
V_5=\frac{1}{\sqrt 2}(v_2+v_5)&$X$&$X$&$X$&$X$&$X$&$X$\\  \hline
V_6=\frac{1}{\sqrt 2}(v_3+v_9)&$X$&$X$&$X$&X&$X$&$X$\\  \hline 
V_7=\frac{1}{\sqrt 2}(v_4+v_{13})&$X$&$X$&$X$&$X$&$X$&$X$\\ \hline 
V_8=\frac{1}{\sqrt 2}(v_7+v_{10})&$X$&$X$&$X$&&$$&\\  \hline 
V_9=\frac{1}{\sqrt 2}(v_7-v_{10})&&&$$&$X$&$X$&$X$\\  \hline
V_{10}=\frac{1}{\sqrt 2}(v_8+v_{14})&$X$&$X$&$X$&&$$&\\  \hline 
V_{11}=\frac{1}{\sqrt 2}(v_8-v_{14})&&$$&&$X$&$X$&$X$\\ \hline 
V_{12}=\frac{1}{\sqrt 2}(v_{12}+v_{15})&$X$&$X$&&$$&$X$&\\  \hline 
V_{13}=\frac{1}{\sqrt 2}(v_{12}-v_{15})&&&$X$&$X$&$$&$X$\\  \hline
\end{array}&
\eea
{\bf Table 1:}  Spanning vectors of the standard finite dimensional  $2$-particle  Hilbert spaces of the $4\times 4$ matrix oscillator. The first four columns correspond to supermatrices:
${\boldsymbol{(4|0)}}$, i.e. the bosonic case, ${\boldsymbol{(3|1)}}$, ${\boldsymbol{(2|2)}}$, i.e. the supersymmetric case,  and  ${\boldsymbol{(1|3)}}$.  The last two columns present the \zzg Hilbert
spaces for parafermions (${{{\mathbb Z}_2^2}}${\bf -PF}) and parabosons (${\mathbb Z}_2^2${\bf -PB}). The ``X" denotes the presence of the vector.\par
~\par
One can observe, in certain cases, the absence of the vectors $V_2,V_3,V_4$. It is a consequence of the Pauli exclusion principle for (para)fermions; this principle is encoded \cite{{top1},{maj1}} in the language of the coproduct.

~\par
The finite-dimensional Hilbert spaces ${\overline{\cal H}}_{k}^{(2)}$ have dimensions $d_k$ given by
\bea
&d_1=d_{6} =10, \qquad d_{2}=9, \qquad d_{3}=d_5=8, \qquad d_{4}=7.&
\eea
The six $2$-particle Hilbert spaces ${\cal H}_k^{(2)}$ recovered from the standard decompositions  are therefore spanned by the vectors $|m; I\rangle$, with $m=0,1,2,\ldots$, while $I$ is restricted according to
\bea\label{inequivalenthilbert}
{\cal H}^{(2)}_1:&&  ~~~ |m; I\rangle \quad{\textrm{for}}\quad m\in {\mathbb N}_0 \quad{\textrm{and}}\quad I=1,2,3,4,5,6,7,8,10,12;\nonumber\\ 
{\cal H}^{(2)}_{2}:&& ~~~ |m; I\rangle \quad{\textrm{for}}\quad m\in {\mathbb N}_0 \quad{\textrm{and}}\quad I=1,2,3,5,6,7,8,10,12;\nonumber\\
{\cal H}^{(2)}_{3}:&& ~~~ |m; I\rangle \quad{\textrm{for}}\quad m\in {\mathbb N}_0 \quad{\textrm{and}}\quad I=1,2,5,6,7,8,10,13;\nonumber\\
{\cal H}^{(2)}_{4}:&& ~~~ |m; I\rangle \quad{\textrm{for}}\quad m\in {\mathbb N}_0 \quad{\textrm{and}}\quad I=1,5,6,7,9,11,13;\nonumber\\
{\cal H}^{(2)}_{5}:&& ~~~ |m; I\rangle \quad{\textrm{for}}\quad m\in {\mathbb N}_0 \quad{\textrm{and}}\quad I=1,2,5,6,7,9,11,12;\nonumber\\ 
{\cal H}^{(2)}_{6}:&& ~~~ |m; I\rangle \quad{\textrm{for}}\quad m\in {\mathbb N}_0 \quad{\textrm{and}}\quad I=1,2,3,4,5,6,7,9,11,13.
\eea

One should note, see (\ref{nsgradings}),  that the three $2$-particle Hilbert spaces recovered from the non-standard decompositions coincide with the associated standard Hilbert spaces.  The difference is encoded in the modified
Hamiltonian, $H_{osc}^{(2)}\mapsto {\overline H}_{osc}^{(2)}$, with the latter given in (\ref{overlinehamosc}). Therefore, we have 
\bea
&{\cal H}^{(2)}_{2}~~{\textrm{for}}~~ {\boldsymbol{(3|1)_{ns}}},\qquad {\cal H}^{(2)}_{3}~~{\textrm{for}}~~ {\boldsymbol{(2|2)_{ns}}},\qquad {\cal H}^{(2)}_{5}~~{\textrm{for}}~~ 
{~~\boldsymbol{{\mathbb Z}_2^2${\bf -PF}$_{ns}}}.&
\eea
~\par
Any vector $|m; I\rangle$ is an energy eigenstate.\par
For the standard quantizations the energy eigenvalues $E_{m,I}$ are read from
\bea\label{energyeigen}
&H_{osc}^{(2)}|m; I\rangle =E_{m,I}|m; I\rangle,\qquad{\textrm{with}}\qquad
E_{m,I}= m+S_I&\nonumber\\
&(S_1=S_2=S_5=0,\quad S_6=S_7=S_8=S_9=S_{10}=S_{11}=1,\quad S_3=S_4=S_{12}=S_{13}=2).&\nonumber\\&&
\eea

For the non-standard quantizations the energy eigenvalues ${\overline E}_{m,I}$ are read from
\bea\label{energyeigenns}
&{\overline H}_{osc}^{(2)}|m; I\rangle ={\overline E}_{m,I}|m; I\rangle,\qquad{\textrm{with}}\qquad
{\overline E}_{m,I}= m+{\overline S}_I&\nonumber\\
&({\overline S}_1={\overline S}_{7}=0,\quad {\overline S}_{5}={\overline S}_6={\overline S}_{10}={\overline S}_{11}={\overline S}_{12}={\overline S}_{13}=1,\quad {\overline S}_{2}={\overline S}_{3}={\overline S}_{8}=
{\overline S}_9=2).&
\eea

For all quantizations (standard and non-standard) the spectrum of the energy eigenvalues $E_n$ is given by the non-negative integers 0,1,2,\ldots:
\bea
E_n&=& n\in {\mathbb N}_0.
\eea
We now discuss the degeneracy of the energy levels and the inequivalence of the multi-particle quantizations.

\subsection{Degeneracy of the energy levels}

The degeneracy of a energy level depends on the given quantization and is obtained from (\ref{energyeigen},\ref{energyeigenns}). The results are summarized in the table below which presents the nine cases ($1$ to $6$ corresponding to the standard decompositions, $7$, $8$ and $9$ to the non-standard ones). For any given quantization the degeneracy of its energy levels $n=2,3,4,\ldots$ is the same. We have

\bea&\nonumber
\begin{array}{|l|c|c|c|}\hline  &E=0&E=1&E=n\geq 2 \\  \hline
1^\ast $-$~ {\boldsymbol{(4|0)}} &3&7&10\\ \hline
2 ~$-$ ~{\boldsymbol{(3|1)}}&3&7&9\\ \hline
3^\dagger $-$~ {\boldsymbol{(2|2)}}&3&7&8\\ \hline
4 ~$-$ ~{\boldsymbol{(1|3)}}&2&6&7\\ \hline
5^\dagger $-$ ~{\boldsymbol{{\mathbb Z}_2^2${\bf -PF}$$ $}}&3&7&8\\ \hline
6^\ast $-$~ {\boldsymbol{{\mathbb Z}_2^2${\bf -PB}$$ $}}&3&7&10\\ \hline
7 ~$-$ ~{\boldsymbol{(3|1)_{ns}}}&2&6&9\\ \hline
8^\ddagger ~$-$ ~{\boldsymbol{(2|2)_{ns}}}&2&6&8\\ \hline
9^\ddagger ~$-$  ~{\boldsymbol{{\mathbb Z}_2^2${\bf -PF}$_{ns}$ $}}&2&6&8\\ \hline
\end{array}&
\eea
\par
~\par
{\bf Table 2:}  The numbers give the degeneracy of the $2$-particle energy eigenvalues for each one of the nine quantizations of the $4\times 4$ quantum oscillator. Different numbers indicate inequivalent quantizations. The inequivalence of the quantizations $1$ versus $6$, $3$ versus $5$ and $8$ versus $9$ cannot be read from this table; it requires a subtler analysis of other observables.\par
~\par
The final proof of the inequivalence of the nine quantizations is given in Section {\bf 5} with the construction of the
observables discriminating the cases $1$ versus $6$ and $8$ versus $9$. The observables discriminating the cases
$3$ versus $5$ are found in \cite{top1}.

~\par
{\bf  Remark:} due to the coassociativity of the coproduct, see (\ref{coassoc}), nine inequivalent $M$-particle
graded Hilbert spaces are recovered for any integer number $M>1$. The formulas are straightforward generalizations
of the $2$-particle construction. In \cite{top1} inequivalent $3$-particle Hilbert spaces were presented for the supersymmetric and (standard) parafermionic gradings. 

\section{Discriminating ${\boldsymbol{2}}$-particle observables}

In (\ref{inequivalenthilbert}) we presented the $2$-particle Hilbert spaces ${\cal H}_k^{(2)}$ (for $ k=1,2,\ldots,6$) which were used to derive the nine (standard and non-standard) quantizations entering Table {\bf 2}. We present here the proof that these nine quantizations are all inequivalent. \\
Since 
the construction of the observables which discriminate the parafermionic case ${\boldsymbol{{\mathbb Z}_2^2}}${\bf -PF} from the supersymmetric case ${\boldsymbol{(2|2)}}$ was given in \cite{top1}, what is left here is to present:\\
{\it i}) observables which discriminate the parabosonic case ${\boldsymbol{{\mathbb Z}_2^2}}${\bf -PB} from the bosonic case ${\boldsymbol{(4|0)}}$,\\
{\it ii}) at least one observable which discriminates the non-standard cases ${\boldsymbol{{\mathbb Z}_2^2}}${\bf -PF}${\boldsymbol{_{ns}}}$ versus ${\boldsymbol{(2|2)}}{\boldsymbol{_{ns}}}$.\par
Let's proceed.
\par

\subsection{Discriminating \zzg parabosons from bosons}

The $2$-particle observables discriminating parabosons from bosons should satisfy the following requirements:\\
~\\
{\it i}) they should apply to both bosonic and parabosonic Hilbert spaces,\\
{\it ii}) they should be hermitian and\\
{\it iii}) they should belong to the $00$-graded sector of the parabosonic theory in order to have real ($00$-graded) eigenvalues.\par
~\par
The following set of $2$-particle observables, constructed in terms of the exchange operators $X_{11}, X_{10}, X_{01}$ introduced in (\ref{xxxmatrices}), satisfy the above three criteria. \par
We have
\bea\label{xstu}
&X_s = X_{10}\otimes X_{10}, \quad X_t = X_{01}\otimes X_{01}, \quad X_u = X_{11}\otimes X_{11}, \quad X_\ast = X_s+X_t+X_u&
\eea
and
\bea\label{ystu}
Y_s&=&({\mathbb I}_4\otimes X_{11}+X_{11}\otimes {\mathbb I}_4)({\mathbb I}_4\otimes X_{10}+X_{10}\otimes {\mathbb I}_4)({\mathbb I}_4\otimes X_{01}+X_{01}\otimes {\mathbb I}_4)+\nonumber\\
&&({\mathbb I}_4\otimes X_{01}+X_{01}\otimes {\mathbb I}_4)({\mathbb I}_4\otimes X_{10}+X_{10}\otimes {\mathbb I}_4)({\mathbb I}_4\otimes X_{11}+X_{11}\otimes {\mathbb I}_4),\nonumber\\
Y_t&=&({\mathbb I}_4\otimes X_{10}+X_{10}\otimes {\mathbb I}_4)({\mathbb I}_4\otimes X_{01}+X_{01}\otimes {\mathbb I}_4)({\mathbb I}_4\otimes X_{11}+X_{11}\otimes {\mathbb I}_4)+\nonumber\\
&&({\mathbb I}_4\otimes X_{11}+X_{11}\otimes {\mathbb I}_4)({\mathbb I}_4\otimes X_{01}+X_{01}\otimes {\mathbb I}_4)({\mathbb I}_4\otimes X_{10}+X_{10}\otimes {\mathbb I}_4),\nonumber\\
Y_u&=&({\mathbb I}_4\otimes X_{01}+X_{01}\otimes {\mathbb I}_4)({\mathbb I}_4\otimes X_{11}+X_{11}\otimes {\mathbb I}_4)({\mathbb I}_4\otimes X_{10}+X_{10}\otimes {\mathbb I}_4)+\nonumber\\
&&({\mathbb I}_4\otimes X_{10}+X_{10}\otimes {\mathbb I}_4)({\mathbb I}_4\otimes X_{11}+X_{11}\otimes {\mathbb I}_4)({\mathbb I}_4\otimes X_{01}+X_{01}\otimes {\mathbb I}_4),\nonumber\\
Y_{\ast}&=& Y_s+Y_t+Y_u.
\eea
Under the ${\bf S}_3$ permutations which interchange the  parabosonic sectors $11,10,01$, the operators $X_s, X_t$ ($Y_s,Y_t$) are mapped into $X_u$ ($Y_u$), while
$X_\ast$ and $Y_\ast$ are ${\bf S}_3$-invariant. Without loss of generality we can therefore consider the
four operators $X_u, X_\ast, Y_u, Y_\ast$. Their $16\times 16$ matrix representations are given in Appendix {\bf B}.\par
For the purpose of making easier the comparison of the bosonic versus parabosonic Hilbert spaces it is convenient to rename the respective vectors $V_I$ entering Table {\bf 1}.\par
They will be expressed in terms of a sign $\varepsilon$ ($\varepsilon=+1$ for bosons, $\varepsilon=-1$ for parabosons); the corresponding finite-dimensional Hilbert spaces will be denoted as ${{\overline\cal H}}_\varepsilon^{(2)}$. We set
\bea\begin{array}{lll}
U_{00,A} = v_1,\qquad &&\\
U_{00,B} = v_6,\qquad &U_{11}=\frac{1}{\sqrt 2}(v_2+v_5),\qquad&W_{11,\varepsilon} =\frac{1}{\sqrt 2}(v_{12}+\varepsilon v_{15}),\\
U_{00,C} = v_{11},\qquad&U_{10}=\frac{1}{\sqrt 2}(v_3+v_9),\qquad&W_{10,\varepsilon}=\frac{1}{\sqrt 2}(v_8+\varepsilon v_{14}), \\
U_{00,D} = v_{16},\qquad  &U_{01}=\frac{1}{\sqrt 2}(v_4+v_{13}),\qquad&W_{01,\varepsilon} =\frac{1}{\sqrt 2}(v_{7}+\varepsilon v_{10}).
\end{array}
\eea
The suffix denotes the ${\mathbb Z}_2\times{\mathbb Z}_2$-grading of the vector in the parabosonic case.\par
The difference between the two Hilbert spaces should be spotted by measuring the subspaces spanned by
 $W_{11,\varepsilon},W_{10,\varepsilon},W_{01,\varepsilon}$ and should appear as $\varepsilon$-dependent eigenvalues.
\par
The eigenvectors of $X_u$ with nonvanishing eigenvalues are $U_\pm$ and $W_{11,\varepsilon}$:
\bea\label{xueigen}
X_uU_\pm = \pm U_\pm \quad {\textrm{for}}\quad U_\pm =U_{00,C}\pm U_{00,D}, && X_uW_{11,\varepsilon}=\varepsilon W_{11,\varepsilon}.
\eea
The eigenvectors of $X_\ast$ with their respective nonvanishing eigenvalues are
\bea\label{xasteigen}\begin{array}{ll}
X_{\ast} (U_{00,B}-U_{00,C})= -(U_{00,B}-U_{00,C}), \qquad& X_{\ast} W_{11,\varepsilon}= \varepsilon W_{11,\varepsilon}\\
X_{\ast} (U_{00,C}-U_{00,D})= -(U_{00,C}-U_{00,D}),\qquad & X_{\ast} W_{10,\varepsilon}= \varepsilon W_{10,\varepsilon}\\
X_{\ast} (U_{00,D}-U_{00,B})= -(U_{00,D}-U_{00,B}), \qquad& X_{\ast} W_{01,\varepsilon}= \varepsilon W_{01,\varepsilon}.
\end{array}
\eea
The presence of the $\varepsilon$ eigenvalues in (\ref{xueigen},\ref{xasteigen}) proves that, by performing $X_u, X_{\ast}$ measurements, one can determine whether a system under consideration is composed by ordinary bosons or by ${\mathbb Z}_2\times{\mathbb Z}_2$-graded parabosons.\par
A basis of eigenvectors with respective eigenvalues for $Y_\ast$ is given by
\bea
Y_\ast U_{00,A} &=& 0,\nonumber\\
Y_\ast (U_{00,B}+U_{00,C}+U_{00,D}) &=&(12+4\varepsilon) (U_{00,B}+U_{00,C}+U_{00,D}),\nonumber\\
Y_{\ast}(U_{00,B}-U_{00,C})& =& -2\varepsilon (U_{00,B}-U_{00,C}), \nonumber\\
Y_{\ast}(U_{00,B}-U_{00,D})& =& -2\varepsilon (U_{00,B}-U_{00,D}),\nonumber\\
Y_{\ast} U_{11}
&=& 2U_{11},\nonumber\\
Y_{\ast} U_{10}
&=& 2U_{10},\nonumber\\
Y_{\ast} U_{01}
&=& 2U_{01},\nonumber\\
Y_{\ast} W_{11,\varepsilon}
&=& (6+4\varepsilon)W_{11,\varepsilon},\nonumber\\
Y_{\ast} W_{10,\varepsilon}
&=& (6+4\varepsilon)W_{10,\varepsilon},\nonumber\\
Y_{\ast} W_{01,\varepsilon}
&=& (6+4\varepsilon)W_{01,\varepsilon}.
\eea
Unlike $X_\ast$, the operator $Y_\ast$ is $\varepsilon$-dependent, see formula (\ref{appb2}), due to the braiding properties of the tensor products in (\ref{ystu}). This observation explains the presence of the $\varepsilon$ sign
in the eigenvalues obtained from the $U_{00,\bullet}$ vectors.\par
The $Y_u$ eigenvectors and eigenvalues are
\bea
Y_u U_{00,A} &=& 0,\nonumber\\
Y_u U_{11}
&=& 2U_{11},\nonumber\\
Y_{u} U_{10}
&=& 0,\nonumber\\
Y_{u} U_{01}
&=& 0,\nonumber\\
Y_{u} W_{11,\varepsilon}
&=& 2 W_{11,\varepsilon},\nonumber\\
Y_{u} W_{10,\varepsilon}
&=& (2+2\varepsilon)W_{10,\varepsilon},\nonumber\\
Y_{u} W_{01,\varepsilon}
&=& (2+2\varepsilon)W_{01,\varepsilon},\nonumber\\
Y_{u} (U_{00,C}-U_{00,D})
&=& -2\varepsilon (U_{00,C}-U_{00,D})
\eea
and, for $\varepsilon=1$,
\bea
Y_u  (U_{00,B}+\frac{1}{2}U_{00,C}+\frac{1}{2}U_{00,D}) &=& 6(U_{00,B}+\frac{1}{2}U_{00,C}+\frac{1}{2}U_{00,D}),\nonumber\\
Y_u (U_{00,B}-U_{00,C}-U_{00,D})&=&0,
\eea
while, for $\varepsilon =-1$, one has
{\small{\bea
Y_u\left(U_{00,B}+\frac{1}{4}(-3\pm\sqrt{17})(U_{00,C}+U_{00,D})\right)&=& (1\pm\sqrt{17})\left(U_{00,B}+\frac{1}{4}(-3\pm\sqrt{17})(U_{00,C}+U_{00,D})\right).\nonumber\\&&
\eea}}

\subsection{Discriminating  two non-standard quantizations}

The matrix operator $X_u$ is an observable for both Hilbert spaces giving the non-standard cases  ${\boldsymbol{{\mathbb Z}_2^2}}${\bf -PF}${\boldsymbol{_{ns}}}$ and ${\boldsymbol{(2|2)}}{\boldsymbol{_{ns}}}$.
We rename the vectors entering the finite-dimensional Hilbert spaces as
\bea
\begin{array}{ccc}
&{\overline V}_1= v_1,~~~~~~~~~~~~&~~~{\overline V}_2=v_6,~~~~~~~~~~~~~ \\
{\overline V}_3=\frac{1}{\sqrt{2}}( v_2+v_5),~~~&{\overline V}_4=\frac{1}{\sqrt{2}}( v_3+v_9),&~~~{\overline V}_5=\frac{1}{\sqrt{2}}( v_4+v_{13}),\\
~{\overline V}_{6,\varepsilon}=\frac{1}{\sqrt{2}}( v_7+\delta v_{10}),~~~&~{\overline V}_{7,\delta}=\frac{1}{\sqrt{2}}( v_8+\delta v_{14}),&~~~~{\overline V}_{8,\delta}=\frac{1}{\sqrt{2}}( v_{12}-\delta v_{15}) .\end{array}
\eea
The sign $\delta=\pm 1$ corresponds to the ${\boldsymbol{{\mathbb Z}_2^2}}${\bf -PF}${\boldsymbol{_{ns}}}$
case for $\delta=-1$ and to the ${\boldsymbol{(2|2)}}{\boldsymbol{_{ns}}}$ case for $\delta=1$.\par
The $X_u$ eigenvalues are read from
\bea
X_u{\overline V}_J~~&=&0 \quad ({\textrm{for}}~~ J=1,2,3,4,5), \nonumber\\
X_u {\overline V}_{6,\delta}&=&0,\nonumber\\
X_u {\overline V}_{7,\delta}&=&0,\nonumber\\
X_u {\overline V}_{8,\delta}&=&\delta {\overline V}_{8,\delta}.
\eea
Due to the presence of the $\delta$ eigenvalue in the last equation, a measurement of $X_u$ allows to discriminate
the two non-standard cases.\par
This concludes the proof of the inequivalence of the nine quantizations presented in Table {\bf 2}.

\section{Conclusions}

This paper presents the $9$ inequivalent multi-particle quantizations of the $4\times 4$ matrix oscillator given in (\ref{hamosc}). Each quantization is recovered from different ${\mathbb Z}_2$- and ${\mathbb Z}_2\times{\mathbb Z}_2$- gradings (and associated statistics) which are consistently imposed on the component particles. $6$ of the quantizations are obtained from the {standard} block-decompositions of supermatrices, $3$ of them from the {non-standard} ones.\par
This analysis completes the multi-particle quantizations discussed in \cite{top1} for just three cases
(bosonic, supersymmetric and the standard version of \zzg parafermions). \par
The extra quantizations presented in the paper include, in particular, a non-standard version of \zzg parafermions and the \zzg parabosonic statistics induced by \zzg Lie {\it algebras}; unlike the
parafermionic statistics induced by \zzg Lie {\it superalgebras}, see 
\cite{{kada1},{kaha1},{kaha2},{kan1}, {tol1}, {stvj1}, {top1}}, this parastatistics has not been previously considered in the literature.\par
Furthermore, we showed that suitable measurements of observables allow to distinguish if a multi-particle system
is composed by \zzg parabosons or by ordinary bosons. This result gives to the notion of \zzg parabosons a legitimate status in physics, proving that it is not just a mathematical artifact void of physically measurable consequences.\par
A next step, in this line of research, would involve the construction of phenomenological \zzg parabosonic models
which could be put to experimental test. A possible scenario could apply to emergent particles in condensed matter.
Concerning model-building, a general framework to construct \zzg parafermionic models was presented in
\cite{akt1} for the classical case and \cite{akt2} for the quantum case. As pointed out in \cite{kuto1}, an extension of the method allows to derive \zzg parabosonic models.\par
On a separate line, the inequivalent multi-particle quantizations induced by gradings shed some light on open
issues regarding the quantization, as discussed in \cite{tod1} both from a historical and an actual perspective. 
\par
~\\

  {\renewcommand{\theequation}{A.\arabic{equation}}
\setcounter{equation}{0}
  {\renewcommand{\theequation}{A.\arabic{equation}}
  \setcounter{equation}{0}  

{\Large{\bf{Appendix A: Relevant formulas for graded (super)algebras}}}\par
~\\
In order to make the paper self-consistent we collect the relevant formulas concerning 
\bea\label{threecases}
{\textrm{{\it ~~~ i})}} && {\textrm{${\mathbb Z}_2$-graded Lie superalgebras,}} \nonumber\\
{\textrm{{\it ~~ ii})}} &&{\textrm{\zzg Lie superalgebras and}}\nonumber\\
{\textrm{{\it ~ iii})}} &&{\textrm{\zzg Lie algebras. }}
\eea
As recalled in the text they induce inequivalent multi-particle quantizations of the $4\times 4$ matrix harmonic oscillator (\ref{hamosc}).
Following \cite{kuto1} and \cite{top1} we use a compact notation  to describe, at once, the three cases. Ordinary
(bosonic) Lie algebras can be assumed to be ${\mathbb Z}_2$-graded Lie superalgebras with empty odd (fermionic) sector. This allows, e. g., to identify the bosonic case listed in Table {\bf 1} with  $(4|0)$ supermatrices.\par
Depending on the case under consideration, a graded algebra ${\mathfrak g}$ is decomposed into 
\bea\label{A1}
i)~~~~~&:&{\mathfrak g} = {\mathfrak g}_0\oplus {\mathfrak g}_1,\nonumber\\
ii) ~{\textrm{and}}~ iii)&:& {\mathfrak g} = {\mathfrak g}_{00}\oplus {\mathfrak g}_{01}\oplus {\mathfrak g}_{10}\oplus {\mathfrak g}_{11}.\eea
The even ($0$) and odd ($1$) generators in {\it i}) are bosonic (fermionic). The four sectors of {\it ii}) and {\it iii}) are described by $2$ bits. The grading of a generator in {\it i}) is given by ${\vec \alpha}\equiv \alpha\in\{0,1\}$. The grading of a generator in
{\it ii}) and {\it iii}) is given by the pair ${\vec \alpha}=(\alpha_1,\alpha_2)$, with $\alpha_{1,2}\in\{0,1\}$.\par
Three respective inner products, with addition ${\textrm{mod}}~2$, are defined:
\bea\label{innerproducts}
~ {i})~: ~~ {\vec\alpha}\cdot{\vec\beta} &:=& \alpha\beta\in \{0,1\},\nonumber\\
~ {ii})~: ~~ {\vec\alpha}\cdot{\vec\beta} &:=& \alpha_1\beta_1+\alpha_2\beta_2\in \{0,1\},\nonumber\\
~ {iii})~:~~  {\vec\alpha}\cdot{\vec\beta} &:=& \alpha_1\beta_2-\alpha_2\beta_1\in \{0,1\}.
\eea
The graded algebra ${\mathfrak g}$ is endowed with the operation $(\cdot,\cdot):{\mathfrak{g}}\times{\mathfrak g}\rightarrow{\mathfrak g}$. \par
Let $a,b,c\in{\mathfrak{g}}$ be three generators whose respective gradings
are ${\vec{\alpha}}, {\vec{\beta}},{\vec{\gamma}}$. The bracket $(\cdot,\cdot)$ is defined as
\bea\label{roundbracket}
(a,b)&:=& ab -(-1)^{{\vec\alpha}\cdot{\vec \beta}}ba,
\eea
resulting in either commutators or anticommutators.\par
The operation satisfies the graded Jacobi identities
\bea
\label{gradedjac}
 (-1)^{\vec{\gamma}\cdot\vec{\alpha}}(a,(b,c))+
 (-1)^{\vec{\alpha}\cdot\vec{\beta}}(b,(c,a))+
 (-1)^{\vec{\beta}\cdot\vec{\gamma}}(c,(a,b))&=&0.
\eea

The grading $\deg[(a,b)]$ of the generator $(a,b)$ is the ${\textrm{mod}}~2$ sum
\bea\label{gradingcomp}
\deg[(a,b)]&=& {\vec{\alpha}}+{\vec{\beta}}.
\eea

{\it Remark:} in the \zzg superalgebra case {\it ii}) the only sectors which are on equal footing and can be interchanged are $10$ and $01$. In the \zzg algebra case {\it iii}) the three sectors $11,10,01$ are on equal footing and can be interchanged. In Sections {\bf 4} and {\bf 5} we made use of this observation. \par
~\par
A graded algebra ${\mathfrak{g}}$ is represented on a graded vector space ${\cal V}$ such that
\bea\label{gradedvector}
{\it i}): ~~{\cal V}={\cal V}_0\oplus {\cal V}_1 ; &\quad& {\it ii})~{\textrm{and}}~{\it iii}):  ~~
{\cal V} = {\cal V}_{00}\oplus {\cal V}_{01}\oplus {\cal V}_{10}\oplus {\cal V}_{11}.
\eea
The grading of a vector $v\in {\cal V}$ is denoted with ${\vec{\nu}}$. Depending on the case, it is either ${\vec{\nu}}\equiv \nu\in\{0,1\}$ or ${\vec{\nu}}=(\nu_1,\nu_2)$ with $\nu_{1,2}\in\{0,1\}$. A generator $a\in{\mathfrak{g}}$ (of grading ${\vec{\alpha}}$)  is represented by the operator ${\widehat{a}}\in End({\cal V})$. The compatibility of the gradings requires that the grading ${\vec{\nu}}'$ of the transformed vector
$v'=av\in {\cal V}$ is ${\vec{\nu}}'={\vec{\alpha}+{\vec\nu}}$. The sum is taken mod $2$.
\par
~\par
Without loss of generality, see the discussion in Section {\bf 4}, we can assume the ${\mathbb Z}_2$-graded matrices
to be split into block-diagonal even and odd sectors. For the \zzg (super)algebras, also without loss of generality,
the $4\times 4$ graded matrices can be decomposed according to
{\small{\bea\label{gradedmatrices}
M_{00}=\left(\begin{array}{cccc} m_1&0&0&0\\0&m_2&0&0\\0&0&m_3&0\\0&0&0&m_4\end{array}\right)\in {\mathfrak{g}}_{00},&&M_{11}=\left(\begin{array}{cccc} 0&m_5&0&0\\m_6&0&0&0\\0&0&0&m_7\\0&0&m_8&0\end{array}\right)\in {\mathfrak{g}}_{11},\nonumber\\
M_{10}=\left(\begin{array}{cccc} 0&0&m_9&0\\0&0&0&m_{10}\\m_{11}&0&0&0\\0&m_{12}&0&0\end{array}\right)\in {\mathfrak{g}}_{10},&&M_{01}=\left(\begin{array}{cccc} 0&0&0&m_{13}\\0&0&m_{14}&0\\0&m_{15}&0&0\\m_{16}&0&0&0\end{array}\right)\in {\mathfrak{g}}_{01},\nonumber\\&&
\eea
}}
where the entries $m_1,m_2,\ldots,m_{16}$ are either constant numbers or, as in (\ref{qqz}), operators.\par
By assuming this convention, the graded vector space ${\cal V}$ is decomposed according to
{\small{\bea\label{gradedvectors}
&v_{00}=\left(\begin{array}{c} v\\0\\0\\0\end{array}\right)\in {\cal V}_{00}, \quad v_{11}=\left(\begin{array}{c} 0\\v\\0\\0\end{array}\right)\in {\cal V}_{11},\quad v_{10}=\left(\begin{array}{c} 0\\0\\v\\0\end{array}\right)\in {\cal V}_{10},\quad
v_{01}=\left(\begin{array}{c} 0\\0\\0\\v\end{array}\right)\in {\cal V}_{01}.
&\nonumber\\
&&
\eea
}}
The construction of the multi-particle states is based, see \cite{{top1},{maj1}}, on the notions of coproduct and  braided tensor product as applied in the context of Hopf algebras. For all three cases {\it i}), {\it ii}), {\it iii}) in (\ref{threecases}) the Universal Enveloping Algebra $U\equiv {\cal U}({\mathfrak{g}})$ of a graded algebra ${\mathfrak{g}}$  is endowed with a Hopf algebra structure. Definition and properties of Hopf algebras can be found in \cite{maj1}. We limit here to recall the properties that we have used in the main text. \par
The coproduct $\Delta$ is a map 
\bea\label{coproduct}
\Delta &:& U\rightarrow U\otimes U
\eea 
which satisfies the coassociativity property
\bea\label{coassoc}
   (\Delta\otimes id)\Delta(U)&=&(id\otimes \Delta)\Delta(U).
\eea
The action $\Delta(u)$ of the coproduct  on a generic element $u\in U$ can be recovered from the action on the identity ${\bf 1}\in {\cal U}({\mathfrak{g}})$, the action on a primitive element $g\in{\mathfrak{g}}$ and from the comultiplication.
We have
\bea\label{coproductprop}
\Delta({\bf 1})&=&{\bf 1}\otimes{\bf 1},\nonumber\\
\Delta(g) &=& {\bf 1}\otimes g+g\otimes {\bf 1},\nonumber\\
\Delta(u_1u_2) &=& \Delta(u_1)\cdot \Delta(u_2).
\eea 
Concerning the braided tensor product, let $a,b,c,d\in {\mathfrak{g}}$. We assume, as before, the grading of $b,c$
to be respectively given by ${\vec \beta},{\vec\gamma}$. The braiding of two tensor spaces is expressed by
the formula
\bea\label{braidedtensor}
(a\otimes b)\cdot (c\otimes d) &=& (-1)^{{\vec\beta}\cdot{\vec\gamma}} ac\otimes bd.
\eea
For the ${\mathbb Z}_2$- and ${\mathbb Z}_2\times{\mathbb Z}_2$- gradings the braiding corresponds to the $(-1)^{{\vec\beta}\cdot{\vec\gamma}}$ sign. Its expression, depending on case {\it i}), {\it ii}) or {\it iii}), is given in formula (\ref{innerproducts}).

\par
~\par
~\\
  \renewcommand{\theequation}{B.\arabic{equation}}
  \setcounter{equation}{0}  

{\Large{\bf{Appendix B: Representations of ${\boldsymbol{2}}$-particle observables}}}\par
~\par

We present here for completeness the $16\times 16$ constant hermitian matrices which realize the $2$-particle observables $X_u$, $X_{\ast}$, $Y_u$, $Y_{\ast}$
introduced in Section {\bf 5}, formulas (\ref{xstu}) and (\ref{ystu}).
These observables allow to discriminate whether the system under consideration is composed by ordinary bosons or
by \zzg parabosons. \par
We have

{\footnotesize{\bea\label{appb1}
X_u&=&\left(\begin{array}{cccc|cccc|cccc|cccc}
0&0&0&0&0&0&0&0&0&0&0&0&0&0&0&0\\
0&0&0&0&0&0&0&0&0&0&0&0&0&0&0&0\\ 
0&0&0&0&0&0&0&0&0&0&0&0&0&0&0&0\\
0&0&0&0&0&0&0&0&0&0&0&0&0&0&0&0\\ \hline
0&0&0&0&0&0&0&0&0&0&0&0&0&0&0&0\\
0&0&0&0&0&0&0&0&0&0&0&0&0&0&0&0\\ 
0&0&0&0&0&0&0&0&0&0&0&0&0&0&0&0\\
0&0&0&0&0&0&0&0&0&0&0&0&0&0&0&0\\\hline
0&0&0&0&0&0&0&0&0&0&0&0&0&0&0&0\\
0&0&0&0&0&0&0&0&0&0&0&0&0&0&0&0\\ 
0&0&0&0&0&0&0&0&0&0&0&0&0&0&0&{\bf 1}\\
0&0&0&0&0&0&0&0&0&0&0&0&0&0&{\bf 1}&0\\ \hline
0&0&0&0&0&0&0&0&0&0&0&0&0&0&0&0\\
0&0&0&0&0&0&0&0&0&0&0&0&0&0&0&0\\ 
0&0&0&0&0&0&0&0&0&0&0&{\bf 1}&0&0&0&0\\
0&0&0&0&0&0&0&0&0&0&{\bf 1}&0&0&0&0&0
\end{array}\right),\nonumber\\
X_{\ast}&=&\left(\begin{array}{cccc|cccc|cccc|cccc}
0&0&0&0&0&0&0&0&0&0&0&0&0&0&0&0\\
0&0&0&0&0&0&0&0&0&0&0&0&0&0&0&0\\ 
0&0&0&0&0&0&0&0&0&0&0&0&0&0&0&0\\
0&0&0&0&0&0&0&0&0&0&0&0&0&0&0&0\\ \hline
0&0&0&0&0&0&0&0&0&0&0&0&0&0&0&0\\
0&0&0&0&0&0&0&0&0&0&{\bf 1}&0&0&0&0&{\bf 1}\\ 
0&0&0&0&0&0&0&0&0&{\bf 1}&0&0&0&0&0&0\\
0&0&0&0&0&0&0&0&0&0&0&0&0&{\bf 1}&0&0\\ \hline
0&0&0&0&0&0&0&0&0&0&0&0&0&0&0&0\\
0&0&0&0&0&0&{\bf 1}&0&0&0&0&0&0&0&0&0\\ 
0&0&0&0&0&{\bf 1}&0&0&0&0&0&0&0&0&0&{\bf 1}\\
0&0&0&0&0&0&0&0&0&0&0&0&0&0&{\bf 1}&0\\ \hline
0&0&0&0&0&0&0&0&0&0&0&0&0&0&0&0\\
0&0&0&0&0&0&0&{\bf 1}&0&0&0&0&0&0&0&0\\ 
0&0&0&0&0&0&0&0&0&0&0&{\bf 1}&0&0&0&0\\
0&0&0&0&0&{\bf 1}&0&0&0&0&{\bf 1}&0&0&0&0&0
\end{array}\right)
\eea}}
and
{\footnotesize{\bea\label{appb2}
Y_u&=&\left(\begin{array}{cccc|cccc|cccc|cccc}
0&0&0&0&0&0&0&0&0&0&0&0&0&0&0&0\\
0&{\bf 2}&0&0&0&0&0&0&0&0&0&0&0&0&0&0\\ 
0&0&0&0&0&0&0&0&0&0&0&0&0&0&0&0\\
0&0&0&0&0&0&0&0&0&0&0&0&0&0&0&0\\ \hline
0&0&0&0&{\bf 2}&0&0&0&0&0&0&0&0&0&0&0\\
0&0&0&0&0&{\bf 4}&0&0&0&0&{\bf 2}&0&0&0&0&{\bf 2}\\ 
0&0&0&0&0&0&{\bf 2}&0&0&{\bf 2}&0&0&0&0&0&0\\
0&0&0&0&0&0&0&{\bf 2}&0&0&0&0&0&{\bf 2}&0&0\\ \hline
0&0&0&0&0&0&0&0&0&0&0&0&0&0&0&0\\
0&0&0&0&0&0&{\bf 2}&0&0&{\bf 2}&0&0&0&0&0&0\\ 
0&0&0&0&0&{\bf 2}&0&0&0&0&0&0&0&0&0&{\bf 2{\boldsymbol{\varepsilon}}}\\
0&0&0&0&0&0&0&0&0&0&0&0&0&0&{\bf 2{\boldsymbol{\varepsilon}}}&0\\ \hline
0&0&0&0&0&0&0&0&0&0&0&0&0&0&0&0\\
0&0&0&0&0&0&0&{\bf 2}&0&0&0&0&0&{\bf 2}&0&0\\ 
0&0&0&0&0&0&0&0&0&0&0&{\bf{ 2{ {\boldsymbol{\varepsilon}}}}}&0&0&0&0\\
0&0&0&0&0&{\bf 2}&0&0&0&0&{\bf 2{{\boldsymbol{\varepsilon}}}}&0&0&0&0&0
\end{array}\right),\nonumber\\
Y_{\ast}&=&\left(\begin{array}{cccc|cccc|cccc|cccc}
0&0&0&0&0&0&0&0&0&0&0&0&0&0&0&0\\
0&{\boldsymbol{2}}&0&0&0&0&0&0&0&0&0&0&0&0&0&0\\ 
0&0&{\boldsymbol{2}}&0&0&0&0&0&0&0&0&0&0&0&0&0\\
0&0&0&{\boldsymbol{2}}&0&0&0&0&0&0&0&0&0&0&0&0\\ \hline
0&0&0&0&{\boldsymbol{2}}&0&0&0&0&0&0&0&0&0&0&0\\
0&0&0&0&0&{\boldsymbol{4}}&0&0&0&0&{\boldsymbol{4+2\varepsilon}}&0&0&0&0&{\boldsymbol{4+2\varepsilon}}\\ 
0&0&0&0&0&0&{\boldsymbol{4}}&0&0&{\boldsymbol{4+2\varepsilon}}&0&0&0&0&0&0\\ 
0&0&0&0&0&0&0&{\boldsymbol{4}}&0&0&0&0&0&{\boldsymbol{4+2\varepsilon}}&0&0\\ \hline
0&0&0&0&0&0&0&0&{\boldsymbol{2}}&0&0&0&0&0&0&0\\
0&0&0&0&0&0&{\boldsymbol{4+2\varepsilon}}&0&0&{\boldsymbol{4}}&0&0&0&0&0&0\\ 
0&0&0&0&0&{\boldsymbol{4+2\varepsilon}}&0&0&0&0&{\boldsymbol{4}}&0&0&0&0&{\boldsymbol{4+2\varepsilon}}\\
0&0&0&0&0&0&0&0&0&0&0&{\boldsymbol{4}}&0&0&{\boldsymbol{4+2\varepsilon}}&0\\ \hline
0&0&0&0&0&0&0&0&0&0&0&0&{\boldsymbol{2}}&0&0&0\\
0&0&0&0&0&0&0&{\boldsymbol{4+2\varepsilon}}&0&0&0&0&0&{\boldsymbol{4}}&0&0\\ 
0&0&0&0&0&0&0&0&0&0&0&{\boldsymbol{4+2\varepsilon}}&0&0&{\boldsymbol{4}}&0\\
0&0&0&0&0&{\boldsymbol{4+2\varepsilon}}&0&0&0&0&{\boldsymbol{4+2\varepsilon}}&0&0&0&0&{\boldsymbol{4}}
\end{array}\right).\nonumber\\
&&
\eea}}
The ${\boldsymbol{\varepsilon}} $ sign entering the (\ref{appb2}) matrices takes the value ${\boldsymbol{\varepsilon =+1}}$ in the bosonic case and ${\boldsymbol{\varepsilon =-1}}$ in the \zzg parabosonic case. Unlike $X_u$, $X_{\ast}$, the operators $Y_u$, $Y_{\ast}$ are ${\boldsymbol{\varepsilon}}$-dependent.
\par
~\par
~
\\ {\Large{\bf Acknowledgments}}
{}~\par{}~\\
I am grateful to Zhanna Kuznetsova for pointing out the relevance of \zzg Lie algebras.\\
 This work was supported by CNPq (PQ grant 308095/2017-0).


\begin{thebibliography}{99}
\bibitem{riwy1} V. Rittenberg and D. Wyler, 
{\it Generalized Superalgebras}, 
{Nucl. Phys.} {\bf B 139}, 189 (1978).
\bibitem{riwy2} V. Rittenberg and D. Wyler, 
{\it Sequences of $Z_2\otimes Z_2$ graded Lie algebras and superalgebras}, 
{J. Math. Phys.} {\bf 19}, 2193 (1978).
\bibitem{sch1} M. Scheunert, 
{\it Generalized Lie algebras}, 
{J. Math. Phys.} {\bf 20}, 712 (1979).
\bibitem{akt1} N. Aizawa, Z. Kuznetsova and F. Toppan, {\it ${\mathbb Z}_2\times{\mathbb Z}_2$-graded mechanics: the classical theory}, Eur. J. Phys. {\bf C 80}, 668 (2020); arXiv:2003.06470[hep-th].
\bibitem{kuto1} Z. Kuznetsova and F. Toppan, {\it 
Classification of minimal \zzg Lie (super)algebras and some applications}, arXiv:2103.04385[math-ph].
\bibitem{top1} F. Toppan, {\it 
\zzg parastatistics in multiparticle quantum Hamiltonians}, J. Phys. A: Math. Theor. {\bf 54}, 115203 (2021);
arXiv:2008.11554[hep-th].
\bibitem{maj1}
S. Majid, {\it Foundations of Quantum Group Theory}, Cambridge University Press, Cambridge (1995).
\bibitem{luri1} J. Lukierski and V. Rittenberg, 
{\it Color-De Sitter and Color-Conformal Superalgebras}, 
{Phys. Rev.} {\bf D 18},  385 (1978).
\bibitem{vas1} M. A. Vasiliev, 
{\it de Sitter supergravity with positive cosmological constant and generalized Lie superalgebras}, 
{Class. Quantum Grav.} {\bf 2},  645 (1985).
\bibitem{jyw1} P. D. Jarvis, M. Yang and B. G. Wybourne, 
{\it Generalized quasispin for supergroups},  
{J. Math. Phys.} {\bf 28},  1192 (1987).
\bibitem{zhe1} A. A. Zheltukhin, 
{\it Para-Grassmann extension of the Neveu-Schwartz-Ramond algebra}, 
{Theor. Math. Phys.} {\bf 71},  491  (1987) ({Teor. Mat. Fiz.} {\bf 71},  218 (1987)).
\bibitem{aktt1} N. Aizawa, Z. Kuznetsova, H. Tanaka and F. Toppan, 
{\it$ \mathbb{Z}_2 \times \mathbb{Z}_2$-graded Lie symmetries of the L\'evy-Leblond equations}, 
{Prog. Theor. Exp. Phys.} \textbf{2016}, 123A01 (2016); arXiv:1609.08224[math-ph].
\bibitem{aktt2} N. Aizawa, Z. Kuznetsova, H. Tanaka and F. Toppan, 
{\it Generalized supersymmetry and L\'evy-Leblond equation}, 
in S. Duarte \textit{et al} (eds), {Physical and Mathematical Aspects of Symmetries}, Springer, Cham, p. 79 (2017);
arXiv:1609.08760[math-ph].
\bibitem{bru1} A. J. Bruce, {\it ${\mathbb Z}_2\times{\mathbb Z}_2$-graded supersymmetry: 2-d sigma models},  J. Phys. A: Math. Theor. {\bf 53}, 455201 (2020); arXiv:2006.08169[math-ph].
\bibitem{brdu1} A. J. Bruce and S. Duplij, 
{\it Double-graded supersymmetric quantum mechanics}', {J. Math. Phys.} {\bf 61}, 063503 (2020); arXiv:1904.06975 [math-ph].	
\bibitem{akt2} N. Aizawa, Z. Kuznetsova and F. Toppan, 
{\it ${\mathbb Z}_2\times {\mathbb Z}_2$-graded mechanics: the quantization}, arXiv:2005.10759[hep-th].
\bibitem{aad1} N. Aizawa, K. Amakawa and S. Doi, {\it ${\cal N}$-extension of double-graded supersymmetric and superconformal quantum mechanics}, J. Phys. A: Math. Theor. {\bf 53}, 065205  (2020); arXiv:1905.06548[hep-th].
\bibitem{yaji1} W. Yang and S. Jing, {\it A new kind of graded Lie algebra and parastatistical supersymmetry},
Science in China Series A: Math. {\bf 44}, 1167 (2001); arXiv:math-ph/0212004.
\bibitem{yaji2} S. Jing, W. Yang and P. Li, {\it Graded Lie Algebra Generating of Parastatistical Algebraic Relations},
Comm. in Theor. Phys.  {\bf 36}, 647 (2001); arXiv:math-ph/0212009.
\bibitem{kada1} K. Kanakoglou and C. Daskaloyannis, {\it Mixed Paraparticles, Colors, Braidings and a new class of Realizations for Lie superalgebras}, arXiv:0912.1070[math-ph].
\bibitem{kaha1} K. Kanakoglou and A. Herrera-Aguilar, {\it Ladder Operators, Fock Spaces, Irreducibility and
Group Gradings for the Relative Parabose Set Algebra}, Int. J. Alg. {\bf 5}, 413 (2011); arXiv:1006.4120[math-RT].
\bibitem{kaha2} K. Kanakoglou and A. Herrera Aguilar, {\it Graded Fock-like representations for a system of algebraically interacting paraparticles},  J. of Phys.: Conf. Ser. {\bf 287}, 012037 (2011); arXiv:1105.4819[math-ph].
\bibitem{kan1} K. Kanakoglou, {\it Gradings, Braidings, Representations, Paraparticles: Some Open Problems}, Axioms {\bf 1}, 74 (2012); arXiv:1210.2348[math-ph].
\bibitem{tol1} V. N. Tolstoy, 
{\it Once more on parastatistics}, 
{Phys. Part. Nucl. Lett.} {\bf{11}},  933 (2014); arXiv:1610.01628[math-ph].
\bibitem{stvj1} N. I. Stoilova and J. Van der Jeugt, {\it The ${\mathbb Z}_2\times{\mathbb Z}_2$-graded
Lie superalgebra $pso(2m+1|2n)$ and new parastatistics representations},  J. Phys. {A}: Math. Theor. {\bf 51}, 135201 (2018); arXiv:1711.02136[math-ph].
\bibitem{kac1}  V. G. Kac, {\it Lie Superalgebras}, Adv. in Math. {\bf 26}, 8 (1977).
\bibitem{bru2} A. J. Bruce, {\it On a ${\mathbb Z}_2^n$-Graded Version of Supersymmetry}, Symmetry 
{\bf 11 (1)}, 
116 (2019); arXiv:1812.02943[hep-th].
\bibitem{aad2} N. Aizawa, K. Amakawa and S. Doi, {\it  ${\mathbb Z}_2^n$-Graded extensions of supersymmetric quantum mechanics via Clifford algebras}, J. Math. Phys. {\bf 61}, 052105 (2019); arXiv:1912.11195[math-ph].
\bibitem{doai1} S. Doi and N. Aizawa, {\it ${\mathbb Z}_2^3$-Graded extensions of Lie superalgebras and superconformal quantum mechanics}, arXiv:2103.10638[math-ph].
\bibitem{brdu2} A. J. Bruce and S. Duplij, 
{\it Double-graded quantum superplane}, Rep. Math. Phys. {\bf 86}, 383 (2020);
arXiv:1910.12950[math.QA].
\bibitem{wit1} E. Witten, {\it Constraints on supersymmetry breaking}, Nucl. Phys. {\bf B 202}, 253 (1982).
\bibitem{dggt1} F. Delduc, F. Gieres, S. Gourmelen and S. Theisen, {\it Non-standard matrix formats of Lie superalgebras}, Int. J. Mod. Phys. {\bf A 14}, 4043 (1999); arXiv:math-ph/9901017.
\bibitem{tod1} I. Todorov, {\it Quantization is a mistery}, Bulg. J. Phys. {\bf 39}, 107 (2012); arXiv:1206:3116[math-ph].
\end{thebibliography}
\end{document}